\documentclass[twoside,twocolumn,9pt]{article}
\usepackage{authblk}
\usepackage[super,sort&compress,comma]{natbib}
\usepackage[version=3]{mhchem}
\usepackage[left=1.5cm, right=1.5cm, top=2.0cm, bottom=2.0cm]{geometry}
\usepackage{balance}
\usepackage{times,mathptmx}
\usepackage{sectsty}
\usepackage{graphicx}
\usepackage{lastpage}
\usepackage{float}
\usepackage{fancyhdr}
\usepackage{fnpos}
\usepackage[english]{babel}
\usepackage{array}
\usepackage{droidsans}
\usepackage{charter}
\usepackage[T1]{fontenc}
\usepackage[usenames,dvipsnames]{xcolor}
\usepackage{setspace}
\usepackage[compact]{titlesec}
\usepackage{hyperref}
\usepackage{subfigure}
\usepackage{amssymb}
\usepackage{caption}
\usepackage{xfrac}
\DeclareUnicodeCharacter{2206}{$\Delta$}
\DeclareUnicodeCharacter{03C0}{$\Delta$}
\begin{document}
\title{Accelerating the Discovery of g-C$_3$N$_4$-Supported Single Atom Catalysts for Hydrogen Evolution Reaction: A Combined DFT and Machine Learning Strategy}
\author{M. V. Jyothirmai\textit{$^{a\ddag}$}, D. Roshini\textit{$^{a\ddag}$}, B. Moses Abraham$^a$ and Jayant K. Singh\textit{$^{a,b\ast}$}}
\affil{\textit{$^a$Department of Chemical Engineering, Indian Institute of Technology Kanpur, Kanpur-208016, India. \\
$^b$Prescience Insilico Private Limited, Bangalore 560049, India.\\
$^{\ast}$E-mail: jayantks@iitk.ac.in\\
$^{\ddag~}$Equal contribution.}}

\date{}
\maketitle

{\Large\textbf{ABSTRACT:}}

Two-dimensional materials supported by single atom catalysis (SACs) are foreseen to replace platinum for large-scale industrial scalability of sustainable hydrogen generation. Here, a series of metal (Al, Sc, Ti, V, Cr, Mn, Fe, Ni, Cu, Zn) and non-metal (B, C, N, O, F, Si, P, S, Cl) single atoms embedded on various active sites of g-C$_3$N$_4$ are screened by DFT calculations and six machine learning (ML) algorithms (support vector regression, gradient boosting regression, random forest regression, AdaBoost regression, multilayer perceptron regression, ridge regression). Our results based on formation energy, Gibbs free energy and bandgap analysis demonstrate that the single atoms of B, Mn and Co anchored on g-C$_3$N$_4$ can serve as highly efficient active sites for hydrogen production. The ML model based on support vector regression (SVR) exhibits the best performance to accurately and rapidly predict the Gibbs free energy of hydrogen adsorption ($\Delta$G$_{H}$) for the test set with a lower mean absolute error (MAE) and a high coefficient of determination (R$^2$) of 0.45 and 0.81, respectively. Feature selection based on the SVR model highlights the top five primary features: formation energy, bond length, boiling point, melting point, and valance electron as key descriptors. Overall, the multistep workflow employed through  DFT calculations combined with ML models for efficient screening of potential hydrogen evolution reaction (HER)  from g-C$_3$N$_4$-based single atom catalysis can significantly contribute to the catalyst design and fabrication. \\

\section{INTRODUCTION}
The ever-increasing energy demands due to rapid exhaustion of non-renewable fossil fuels enforce the development of clean and sustainable energy technologies.\cite{Dachang,Cook,Esswein,Lei.Y} In this regard, hydrogen is considered as an emerging alternative to replace currently used fuel sources due to its promising features, including high gravimetric energy density, abundance and environmental friendliness.\cite{Gong,Rajamathi} Over the last few decades, extensive research has been carried out both experimentally and computationally to design and develop suitable catalysts for hydrogen production.\cite{Vennapoosa} However, the conversion and storage of hydrogen is a major concern that has a direct impact on the hydrogen evolution reaction (HER).\cite{Bhoyar,Vennapoosa1} Although conventional catalysts like noble metals, platinum, rhodium, etc, can improve activity by enhancing the efficiency of HER, but the sparsity and high cost limits their commercial usage.  Thus, large-scale practical applications require high activity with low cost. Many efforts have been made to reduce the size of metal catalysts and the utilization of non-noble metals\cite{Gao.F,Liu.J} to develop advanced technologies toward
hydrogen generation for unlocking the challenges in energy,  environment and sustainability.

Single atom catalysts (SACs)\cite{Qiao}  has triggered world-wide research interest and emerged as promising alternatives for the traditional noble metal catalysts, benefiting from their maximized atom utilization, highly exposed active centers, unsaturated coordination and atomic dispersion. \cite{Qiao,Yang.T,Wang.A}  In such catalysts, the active sites are mainly dispersed as  single atoms, sub-nanometer clusters, or both, such as Cu-N@C, Co-N-C, Pd@RCC3, Pd-Ni(OH)$_2$@S-1.\cite{Wu.H,Liu.W,Sun.Q,Yang.X,Fei.H} Cheng et al.\cite{Cheng.N} analyzed the performance of Pt dispersed single-atom catalyst and  their results demonstrate the importance of potential substitutes in enhancing the HER efficiency when compared  to its traditional noble metal counterpart. Ragan et al.\cite{Santillan} studied the stability of single atom Pt on pristine-, defected-, doped-graphene using DFT calculations and concluded that the state of the substrate strongly influences the stability as well as catalytic activity. However, the compatibility of the constitute supporting materials to accommodate the atomically dispersed atoms typically influence the structural stability of the SAC\cite{Yang.C}, where the aggregation of dispersed single atoms into nanoclusters or particles is a major concern\cite{Yang}. This recombination is due to a decrease in particle size of the supported noble metal to a single atom, which drastically increases the surface area and thus leads to a sharp rise in the surface free energy of the noble metal. Thus, the immobilization of single-atom sites on 2D surfaces through strong bonding is highly essential for their usage in various applications.

Extensive studies have been reported on single metal atoms dispersed on two-dimensional (2D) substrates (graphene, transition metal dichalcogenides, graphitic carbon-nitride, hexagonal boron nitride, germanane, silicene, phosphorene, black phosphorus, borophene, MXenes, etc) because of their large surface area and  exceptional catalytic properties\cite{Li.W,Anasori,Chia,Deng, Moses,Malik11}. Among these 2D catalysts, graphitic carbon-nitride (g-C$_3$N$_4$)  has been demonstrated as a potential support to stabilize single-atom sites and high density ultra-small metal clusters due to its  “six-fold cavities” and  homogeneous high-density N atoms, which helps in trapping transition metals.\cite{Yang.C,Gao} In addition, anchoring metal catalysts on g-C$_3$N$_4$ substrate offers several inherent benefits, such as identification of catalytically active sites by providing more precise information, effective accumulation of surface polarization charges on metal atoms, and maintaining metal atoms in their neutral state. Several studies have successfully prepared single Fe, Co, Cu, Ni, Ru, Pt, and Pd atoms supported g-C$_3$N$_4$. Combined with the superior catalytic activity of single-atom metal sites and the photo-response characters of g-C$_3$N$_4$, the SAC@g-C$_3$N$_4$ show promising features in the field of photocatalysis. The loading of a single-atom metal can facilitates the separation and transfer of photocarriers by modifying the electron and energy structure of g-C$_3$N$_4$, thereby enhancing light absorption. For instance, the photocatalytic activity of highly dispersed Pt/g-C$_3$N$_4$ co-catalyst was found to enhance the HER.\cite{Li.X}  However, atoms with high electronegativity may probably deactivate or poison the active centres due to highly oxidized metal single atoms. \cite{Zhou,Zhou.P} Such microscopic observations cannot be addressed through mainstream experimental investigations and thus advanced quantum mechanical calculations are highly required to predict the interactions between the single atoms and the support.

In recent years, data-driven approaches applied to materials design have shown potential to overcome such stumbling blocks. Machine-learning (ML) techniques\cite{Butler,Artrith,Keith} simplify the understanding of complex configurational spaces through readily available physical properties. For example, ML models predicted the identification of aggregation trends compromising single atom stability\cite{Su}, explored the impact of surface modifications on MgO(100) supports\cite{Liu} and predicted the binding energies of SACs on various oxide\cite{Tan}. Inspired by the above-mentioned SAC, g-C$_3$N$_4$ and machine learning, here we explore the influence of single metal (Al, Sc, Ti, V, Cr, Mn, Fe, Ni, Cu, Zn) and  non-metal (B, C, N, O, F, Si, P, S, Cl) embedded in g-C$_3$N$_4$ (SACs@g-C$_3$N$_4$) for enhancing the performance of HER. For this purpose, a three-tier screening based on formation energy, Gibbs free energy and band-gap analysis is employed to down-select the suitable candidates. Also, a robust machine learning model is built based on the data generated from DFT as well as the features of gas-phase atoms present in the chemical composition. As a part of model development, the performance of support vector regression (SVR), gradient boosting regression (GBR), random forest regression (RFR), AdaBoost regression (ABR), multilayer perceptron (MLP) regression, ridge regression (RR) algorithms are compared. In addition, feature importance analysis is used to identify key descriptors for the HER. Overall, these findings furnish a fundamental basis for designing novel g-C$_3$N$_4$-based SACs through the prediction of potential indicators for HER.

\begin{figure*}[t]
\centering
\includegraphics[height =4.3in,width=7in]{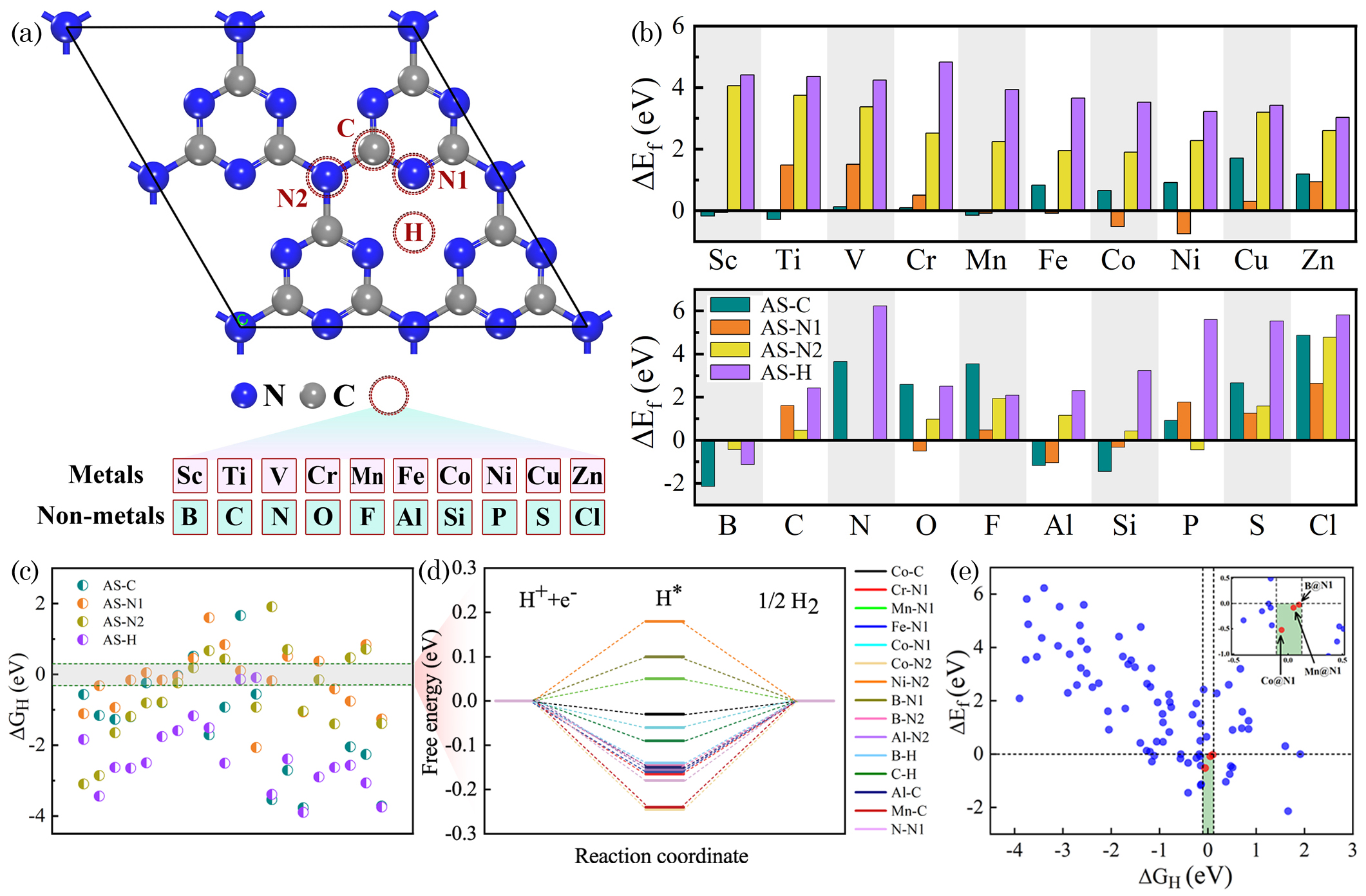}
\caption{(a) The top view of the optimized geometry of g-C$_3$N$_4$ catalyst. The grey and blue color balls represent the C and N  atoms, respectively. The dashed circles with letters indicate various dopant sites: 2-fold coordinated nitrogen with two C atoms (N1), nitrogen connecting with three triazine rings (N2), carbon that bridges three N atoms (C) and center of the four-fold triazine cavity (H) of g-C$_3$N$_4$. (b) The formation energies ($\Delta$E$_F$) of single metal (Al, Sc, Ti, V, Cr, Mn, Fe, Ni, Cu, Zn) and non-metal (B, C, N, O, F, Si, P, S, Cl) atom anchored on g-C$_3$N$_4$. (c) Gibbs free energy ($\Delta$G$_{H}$) distribution of H-adsorption on g-C$_3$N$_4$-based SACs. The black dashed line range represents |$\Delta$G$_H$| < 0.3 eV. (d) Free energy profile of hydrogen evolution for g-C$_3$N$_4$-based SACs in the range of |$\Delta$G$_H$| < 0.3 eV. (e) The stability and performance analysis for HER catalysis using $\Delta$E$_F$ and $\Delta$G$_H$. The candidates (B@N1-site, Mn@N1-site and Co@N1-site) located at the green area (bottom center) are stable with high HER activity.}
\end{figure*}

\section{COMPUTATIONAL DETAILS}
The density functional theory (DFT) calculations were performed using the pseudopotentials and plane-wave basis set via the Vienna ab initio simulation package (VASP) \cite{Kresse,Kresse.G}. The structure relaxation and electron-ion interaction was described using Perdew-Burke-Ernzerhof exchange-correlation functional within the generalized gradient approximation (PBE-GGA) \cite{Perdew} and projector augmented wave (PAW) method \cite{Blochl}, respectively. The Monkhorst-Pack grid \cite{Monkhorst} is used to sample the first Brillouin zone and a K-point mesh of 5x5x1 and 7x7x1 is employed for structural optimization and electronic structure calculations, respectively. The forces are said to be converged upon ionic relaxation when the Hellmann-Feynman forces on each atom between steps were less than 0.01 eV/\AA \space and the convergence criterion for energies between two consecutive SCF cycles was set to 10$^{-5}$ eV. To overcome the disadvantages of standard DFT-PBE calculations in computing an accurate band gap, we employed the Heyd-Scuseria-Ernzerhof (HSE06) hybrid functional \cite{Heyd} with mixing parameters of 0.25.  The long-range interactions were included through a semi-classical dispersion correction scheme (DFT-D3) \cite{Grimme}. To mitigate image interactions used in the plane-wave DFT calculations due to the periodic boundary conditions, a vacuum layer of 15 Å was considered along the direction perpendicular to the surface.

To assess the HER activities of various SACs@g-C$_3$N$_4$, we calculated the hydrogen adsorbed Gibbs free energy ($\Delta$G$_{H}$) using the following equation:
\begin{equation}
\Delta G_{H} = \Delta E_{H} + \Delta E_{ZPE} - T\Delta S_{H}
\end{equation}

where $\Delta$E$_{H}$ corresponds to the H-atom adsorption energy on the surface of SACs@g-C$_3$N$_4$. $\Delta$E$_{ZPE}$ is the change in zero-point energy under harmonic approximation before and after hydrogen adsorption.  $\Delta$S$_{H}$ represents the difference in entropy between the gaseous hydrogen and adsorbed hydrogen. T is the temperature (298.15 K). $\Delta$E$_{H*}$ is computed based on the equation:

\begin{equation}
\Delta E_{H}  = E_{H} - E_{surface} - \frac{1}{2}E_{H_2}
\end{equation}
where E$_{H}$, E$_{surface}$ and E$_{H_2}$ is obtained from DFT calculations by computing the total energies of SACs@g-C$_3$N$_4$ catalysts with, without adsorbed H atom and isolated H$_2$ gas molecule, respectively.

\subsection{Machine Learning Methodology}
An open-source Python distribution platform is used to perform machine learning strategies with scikit-learn libraries\cite{Scikit}. The ML approach in the present work contains two parts: feature engineering, followed by model prediction. Initially, 28 primary features that may influence the Gibbs free energy ($\Delta$G$_{H}$) are considered. The data required for conducting such studies were generated from the features of gas-phase atoms that are used to build g-C$_3$N$_4$-based SAC's and their corresponding pristine properties (see Table S1). Subsequently, to process every type of feature, we performed feature engineering through statistical functions (FESF) via mean, standard deviation and variance, which increased the number of primary features from 28 to 60. Finally, the leave-one-out approach\cite{OrialP} is employed to remove the unimportant features. These selected features and the computed $\Delta$G$_{H}$ were used as input and output data, respectively. All the g-C$_3$N$_4$-based SAC's analyzed in the present work were randomly distributed into a training set and test set with a ratio of 9:1. For model prediction, we used six different ML models, such as SVR, GBR, RFR, ABR, MLP and RR, to train the data. The hyperparameter tuning incorporated for the  top selected SVR and MLP  models are presented in Table S2.  The coefficient of determination (R$^2$) and mean absolute error (MAE) are used to evaluate the accuracy of the model.

\begin{figure*}[t]
\centering
\includegraphics[height =3.5in,width=7in]{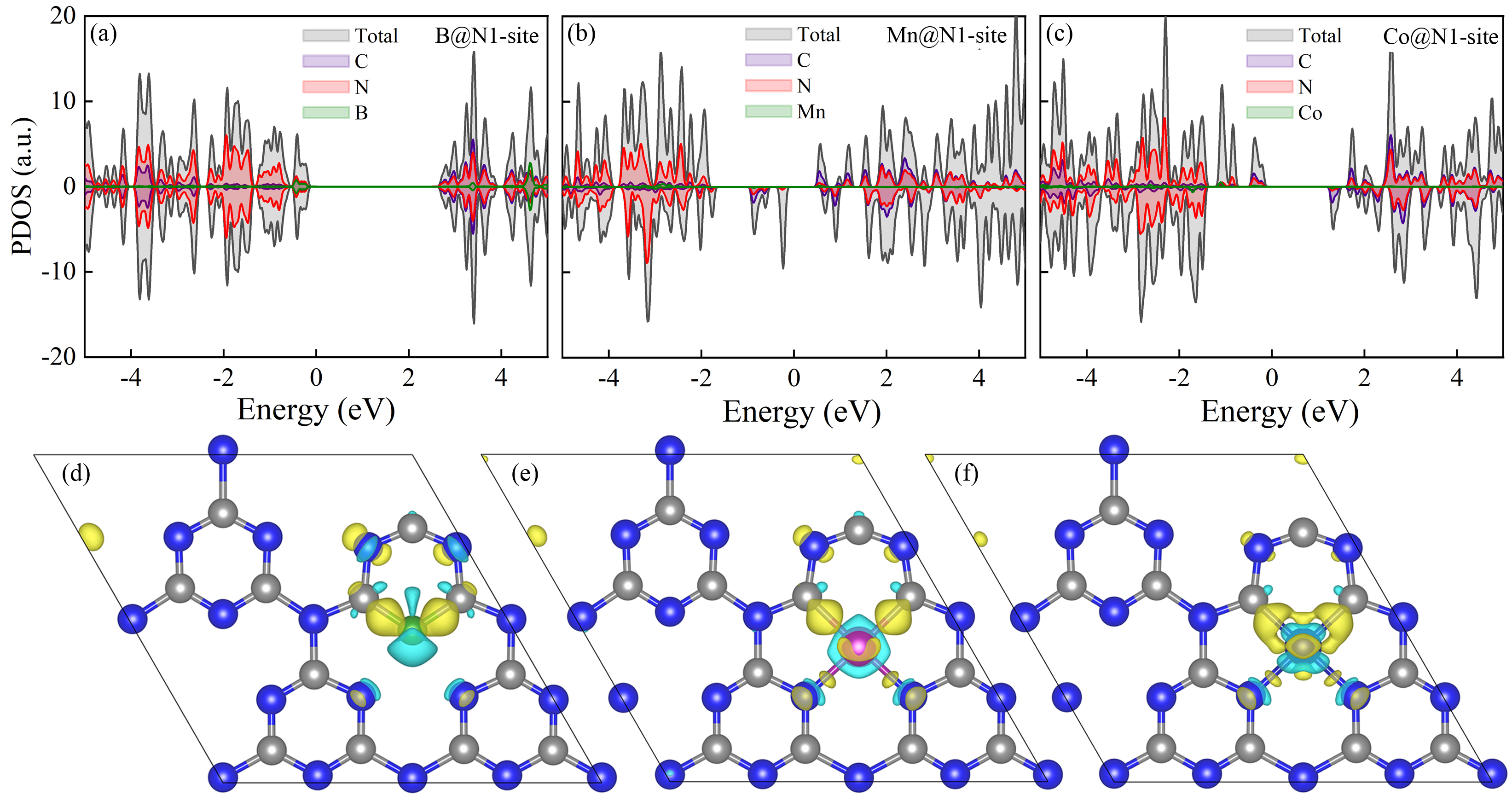}
\caption{The partial density of states and corresponding charge density difference of (a and d) B@N1-site, (b and e), Mn@N1-site  and (c and f) Co@N1-site. The yellow and blue iso-surface indicates electron accumulation and depletion, respectively.}
\end{figure*}

\section{RESULTS AND DISCUSSION}
 To rationally design single metal/non-metal atoms immobilized on the g-C$_3$N$_4$ for HER catalysts,  we employed high throughput screening. Accordingly, we first constructed the  g-C$_3$N$_4$ monolayer by cleaving the corresponding bulk structure. After geometry optimization, g-C$_3$N$_4$ has bond lengths of  1.57 \AA \space (C–N1) and  1.35 \AA \space (C–N2), which are in accord with the reported results. Subsequently, a single metal (Al, Sc, Ti, V, Cr, Mn, Fe, Ni, Cu, Zn) or non-metal (B, C, N, O, F, Si, P, S, Cl) atom is anchored on g-C$_3$N$_4$ for constructing SACs.  As shown in Figure 1a, four different types of doping sites such as 2-fold coordinated nitrogen with two C atoms (N1), nitrogen connecting with three triazine rings (N2), carbon that bridges three N atoms (C) and center of the four-fold triazine cavity (H) of g-C$_3$N$_4$  are considered for doping the selected  metal and non-metal atoms. A basic prerequisite for single atom catalysts embedded in g-C$_3$N$_4$ is to maintain good catalytic activity and high stability in long-term usage. In other words, the doped elements should bind strongly with g-C$_3$N$_4$ to avoid aggregation of atoms. To explore the favourable loading site for single metal/non-metal doping on g-C$_3$N$_4$, the formation energies ($\Delta$E$_F$) is computed using: $\Delta$E$_F$ = E$_{Total}$ - E$_{g-C_3N_4}$ - $\mu(A)$ + $\mu(B)$. Here, E$_{Total}$  is the total energy of metal/non-metal doped g-C$_3$N$_4$ (M/N@g-C$_3$N$_4$), E$_{g-C_3N_4}$ is the total energy of g-C$_3$N$_4$, $\mu$(A) and $\mu$(B) are the chemical potentials of single metal/non-metal and C/N atoms, respectively. According to the definition, the negative formation energy indicates structural stability. As shown in Figure 1b, most of the metal and non-metal doping have positive formation energy, while Sc, Ti, Mn, B, Al and Si at  C-site, Sc, Mn, Fe, Co, Ni, B, O, Al and Si at N1-site, B and P at N2-site and B at H-site demonstrate negative $\Delta$E$_F$, which implied their structural stability and more likely to be synthesized. Particularly, Al@C-site (-1.17 eV) and B@C-site (-2.14 eV) have the lowest formation energy in metal and non-metal cases, respectively, which is favorable  for preventing metal aggregation or being leached.\cite{Xu} The anisotropic nature of formation energies indicates that the loading of single metal/non-metal atoms significantly influences the stability of g-C$_3$N$_4$. Among four doping sites, the atoms embedded in N1 site are preferable when compared to other possible sites. Previous studies demonstrated that the atoms tend to substitute in the edge nitrogen atom of g-C$_3$N$_4$.\cite{Ge.L,Lu.C,Lin.S}, indicating that the N1-site is favorable for fabricating SACs. Typically, a site is called as active for doping single atoms if it contains negative formation energy. Here, N2- and H-sites are not favorable sites for generating stable SAC@g-C$_3$N$_4$. Thus, anchoring appropriate atoms at suitable position are highly required to  maintain favorable stability of g-C$_3$N$_4$.  It is also noteworthy that the bond lengths of SAC@g-C$_3$N$_4$ are longer than that of pure counterparts, which might be due to the difference in electronegativity of C/N and dopants. We notice that the formation energy criteria  eliminates 78\% of the total 80 configurations. Therefore, 18 candidates with negative formation energy are considered as favorable structures.

\begin{figure*}[t]
\centering
\includegraphics[height =4.1in,width=6in]{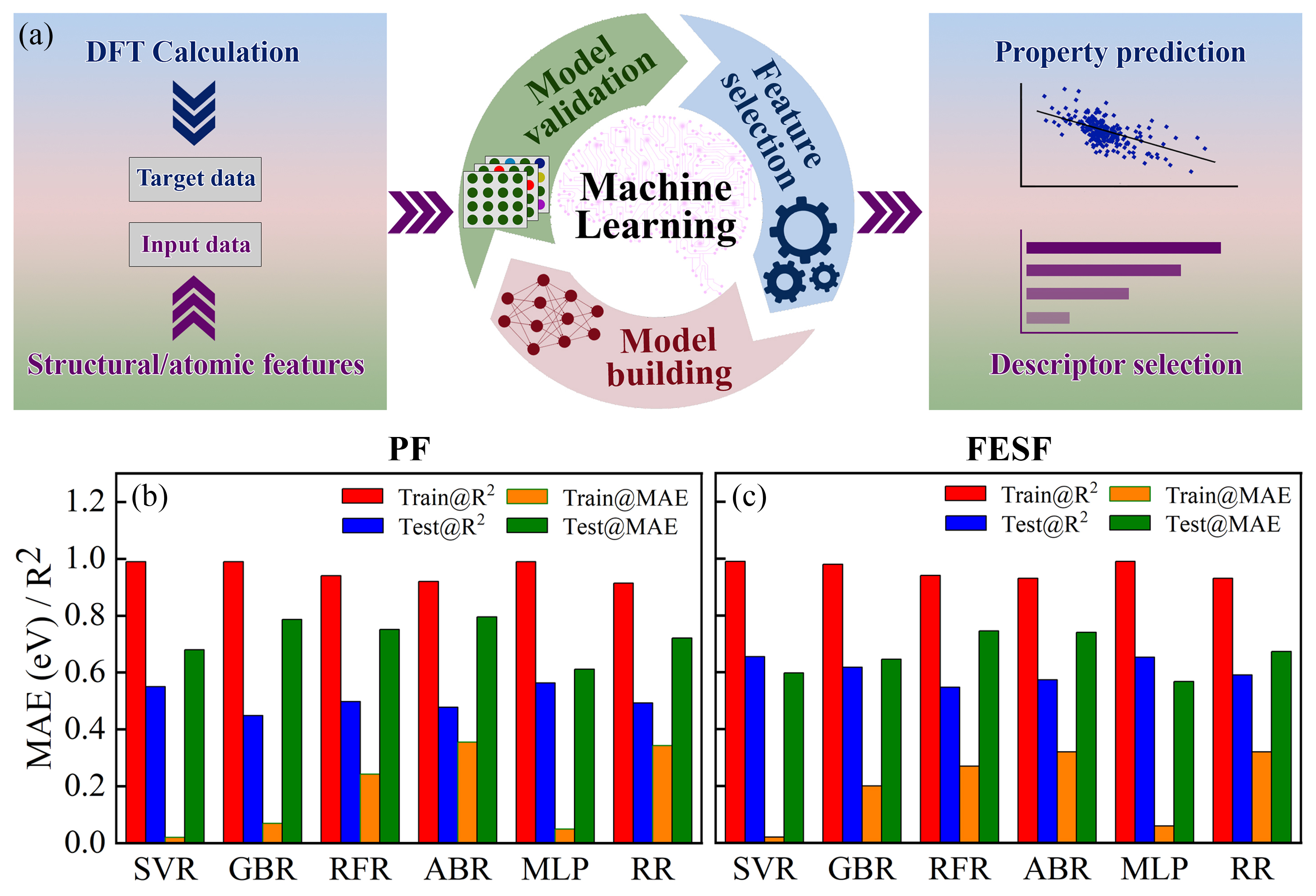}
\caption{(a) Schematic representation demonstrating the process of machine learning. The complete process involves three stages: data generation, model training and test, followed by feature analysis. The coefficient of determination (R$^2$) and mean absolute error (MAE) of SVR, GBR, RFR, ABR, MLPR and RR algorithms using (b) primary features (PF) and (c) feature engineering through statistical functions (FESF). Red, blue, orange, and blue bars represent the training R$^2$, testing R$^2$, training MAE, and testing MAE,
respectively.}
\end{figure*}

 According to the computational hydrogen electrode (CHE) model,\cite{Greeley} a catalyst is said to be a potential candidate for efficient HER activity if it possesses |$\Delta$G$_H$| smaller than 0.3 eV. It should be noted that both absorption and desorption of hydrogen are required for the hydrogen evolution process; either too weak or too strong H-adsorption typically lowers the reaction rate: A highly positive $\Delta$G$_H$ makes the proton bond strongly with the catalyst, while the highly negative $\Delta$G$_H$ tends to have a strong binding with the adsorbed hydrogen atom. Typically, the HER process involves three hydrogen steps: initial step (H$^+$/e$^-$ couples),  intermediate step (H*), and the final product (1/2 H$_2$). The intermediate step, i.e., the $\Delta$G$_H$ for H-adsorption is a universal descriptor for HER activity following either the Heyrovsky or the Volmer-Tafel mechanism.\cite{Hinnemann} Figure 1c,d displays the $\Delta$G$_H$ profile of H-adsorption on single atom-loaded g-C$_3$N$_4$ at various possible active sites. The loading of single metal/non-metal atoms in g-C$_3$N$_4$ led to a drastic change in |$\Delta$G$_H$|, which demonstrates the significance of dopants in tuning the HER activity of g-C$_3$N$_4$ catalyst. In the case of H-site, the highly negative $\Delta$G$_H$ indicates the strong attraction between the adsorbed H* and dopants on SAC@g-C$_3$N$_4$. The strong adsorption  on H-site inhibits the desorption of H-atom from the SAC@g-C$_3$N$_4$ surface. For N1-site, the $\Delta$G$_H$ for most of the cases fall in the optimum range, indicating that their HER performance is higher when compared to other doping sites. Among all, only 15 candidates (6 from N1-site, 3 from C-site, 4 from N2-site and 2 from H-site) from 80 configurations show |$\Delta$G$_H$| values less than 0.3 eV, which indicates the excellent HER catalyst activity. Moreover,  the stability and performance analysis for HER catalysis using $\Delta$E$_F$ and $\Delta$G$_H$ (see Figure 1e) demonstrates 3 potential candidates (B@N1, Mn@N1, Co@N1) with |$\Delta$G$_H$| values less than 0.09 eV may have catalytic activity surpassing that of Pt (111).

\begin{figure*}[t]
\centering
\includegraphics[height =3.8in,width=6in]{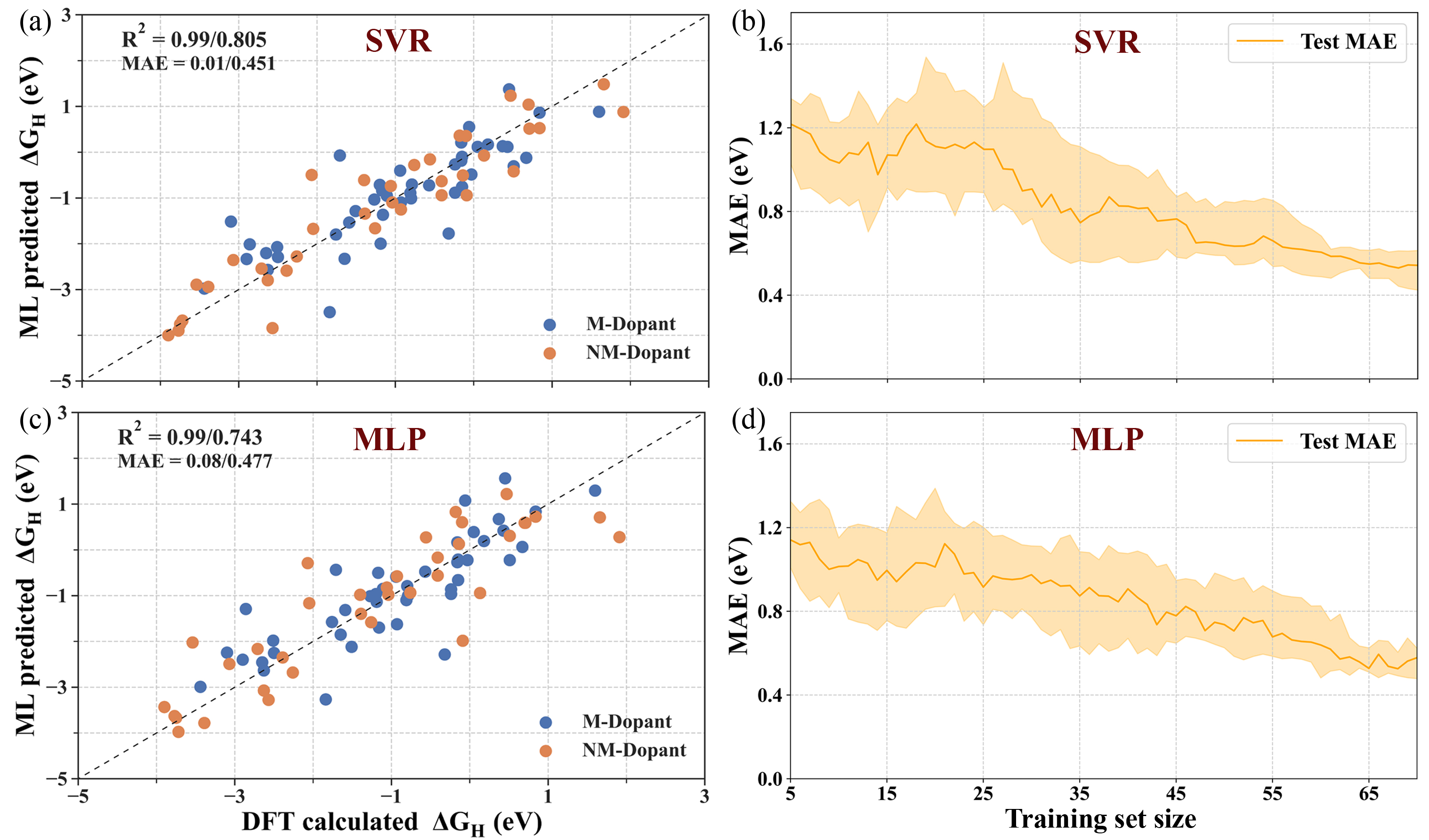}
\caption{Calculated vs predicted $\Delta$G$_H$ and corresponding mean absolute error (MAE) evolution of the test set versus training size  for (a-b) SVR and (c-d) MLP models using FESF with hyperparameter tuning. The blue and orange color circles indicate the single metal (Al, Sc, Ti, V, Cr, Mn, Fe, Ni, Cu, Zn) and non-metal (B, C, N, O, F, Si, P, S, Cl) single atoms embedded on various active sites of g-C$_3$N$_4$ The shaded regions indicate the standard deviation limits.}
\end{figure*}

A strong interaction between the single atom and the support highly influences the electronic structure that eventually regulates the catalytic properties of SACs. To reveal the effect of doping on the band gap of g-C$_3$N$_4$, we analyzed the partial density of states (PDOS) of pure and metal/non-metal embedded in g-C$_3$N$_4$ using the HSE functional. As shown in Figure S1, the pure g-C$_3$N$_4$ is found to be a semiconductor with a band gap of 2.93 eV, which is in agreement with the previous studies. The  PDOS reveals that the conduction band minimum (CBM) of intrinsic g-C$_3$N$_4$ is mainly composed of both C and N atoms, while the valence band maximum (VBM) is dominated by N atom. The PDOS and corresponding charge density difference (CDD) of the down-selected B@N1 Mn@N1 and Co@N1 configurations are shown in Figure 2a-f.  In the case of B@N1, the  spin-up and spin-down components of the electronic states are symmetrically distributed, indicating the non-magnetic nature. The band gap decreases to 2.02 eV after the introduction of boron into g-C$_3$N$_4$ and the electronic states are adjusted towards the low energy region. The shift in the electronic states is not surprising as the electronegativity of boron is smaller than nitrogen, which leads to the creation of valence electrons that occupies a part of the conduction band. Moreover, the orbitals of the B atom merge with the C atom, indicating a solid hybridization between them and thus confirming the strong covalent interaction between boron and carbon atoms. This will help in the catalytic reaction to improve the reactivity and activity of the adsorbates. The Bader charge analysis also demonstrates a notable charge transfer from boron to the g-C$_3$N$_4$ surface. This supports strong covalent atom-substrate interaction due to the stabilization of boron atom on g-C$_3$N$_4$. The charge density difference (CDD) confirmed that there is prominent charge depletion and accumulation around B and its neighboring atoms, respectively. The embedded B atom display loss of electron density, which is in accord with the flow of charge density from boron to the nearest atoms. In the case of Mn and Co at N1-site, we observe an asymmetric distribution of minority and majority spin components, leading to spin polarization in the system due to the introduction of metal dopants. Especially, the VBM of Mn (Co) at N1-site is mainly contributed by a minority (majority) spin channel, while the CBM is dominated by the majority (minority) spin component. The incorporation of Mn and Co atoms introduces  localized impurity levels in the band gap region with spin down and spin up channels. The observed behavior is expected as the dopants influence the C and N atoms of g-C$_3$N$_4$ and thereby leading to strong hybridization. Due to the presence of impurity levels, both the spin-up channel and spin-down channel of Mn (Co) at N1-site are semiconductors with a band gap of 2.24 (1.69) and 0.61 (2.56) eV, respectively. This indicates that the photogenerated electron can comfortably jump from the VB to the impurity state  or from the impurity state to the CB. Under UV–vis light irradiation, such behavior in the electronic properties of SAC@g-C$_3$N$_4$ can significantly enhance the photocatalytic activity.

\section{MACHINE LEARNING DISCOVERY OF CATALYST DESCRIPTORS}

Although the change in $\Delta$G$_H$ can be a quantitative description to a certain degree through a simple linear relationship, however, the correlation between atomic/structural properties and $\Delta$G$_H$ is still uncertain. To address this issue, we introduce machine learning (ML) algorithms to our calculated data for rapid screening of single atom-loaded g-C$_3$N$_4$ and their HER evolution. The schematic representation demonstrating the process of machine learning is shown in Figure 3a. For training a fast and efficient ML model, feature selection analysis is very important for the quantitative discovery of semi-automated correlations in materials science and chemistry. Since the electronic characteristics of catalysts influence the HER activity, we have considered 28 primary features that are related to atomic and structural properties of SACs@g-C$_3$N$_4$. ML algorithms, such as SVR, GBR, RFR, ABR, MLP and RR, were employed to predict the $\Delta$G$_H$. The  DFT-calculated vs model predicted $\Delta$G$_H$ values are shown in Figure S2. Typically,  a higher coefficient of determination (R$^2$) and a lower mean absolute error (MAE) through model training and test are expected to enhance the accuracy of the ML model. The minimum/maximum R$^2$ and  MAE are found to be 0.44/0.56 and 0.61/0.79 for GBR/MLP and MLP/ABR, respectively (see Figure 3b). Clearly, all the studied ML models show poor predicted performance with a low R$^2$ and high MAE values in both training and testing sets. This indicates that the limited input data is unable to predict the comprehensive analysis of HER activity without extensively accounting the impact of each constituent atom in SACs@g-C$_3$N$_4$. In this scenario, statistical functions were employed to process every type of features, such as mean, standard deviation and variance. Accordingly, the application of statistical functions increased the number of primary features from 28 to 60.  As shown in Figure 3c and S3, the training/testing R$^2$ values after engineering the features using statistical functions are found to be 0.99/0.65, 0.98/0.62, 0.94/0.55, 0.93/0.57, 0.99/0.65 and 0.93/0.59 for SVR, GBR, RFR, ABR, MLP and RR, respectively. Combined with MAE, both training and testing sets of SVR and MLP models show better predictive performance towards $\Delta$G$_H$.

Typically, the training process of ML model can be accelerated by using easily available features. Although the time consumed for obtaining these properties is relatively lower than the DFT computed $\Delta$G$_H$, it is important to remove the overlapping features to further accelerate the predicted process. Moreover, the presence of unnecessary features may lead to low training efficiency and high prediction bias.\cite{Jager} Therefore, the feature selection is finalized by reducing input data using the leave-one-out approach. In addition, to evaluate the ability of the model, hyperparameter were tuned via 10-fold cross-validations to maximize the accuracy when compared to its counterpart. Using this parameterization, the SVR and MLP models were evaluated again to predict the performance over the same randomized data as shown in Figure 4a-d. Compared to MLP, the predicted performance of SVR model is significantly improved with the low MAE of 0.01 eV and high R$^2$ of 0.99 during model training and also works well for the test set with the MAE of 0.45 eV and R$^2$ of 0.81 eV. Overall, SVR is demonstrated to be a promising model with stable and reliable predictive performance. In this regard, it is necessary to explore the feature importance of a catalyst to evaluate the inherent correlation between the target property and the primary features, thereby providing key insights into the origin of H-adsorption ability. For the final SVR model with FESF after hyperparameter tuning, the top five key descriptors that highly correlate with the HER activity of SACs@g-C$_3$N$_4$ are formation energy, bond length, valance electron, melting point, and boiling point. Among them, formation energy and bond length are DFT computed descriptors, while valance electron, melting point, and boiling point are elemental properties. Taking the high ranking of these descriptors into account, the accurate regression of the significant SVR-based ML model mainly relies on the structural and atomic features and also reveals the underlying correlation of these descriptors with the catalytic activity.

\section{CONCLUSION}
In summary, the feasibility of anchoring single metal or non-metal atoms on g-C$_3$N$_4$ has been systematically investigated for efficient HER catalysts by combining first-principles DFT calculations and  machine learning strategies. Among the studied 80 configurations of 20 metal and non-metal embedded g-C$_3$N$_4$, only 18 candidates (Sc, Ti, Mn, B, Al and Si at  C-site, Sc, Mn, Fe, Co, Ni, B, O, Al and Si at N1-site, B and P at N2-site and B at H-site) show negative formation energy, which implied their structural stability and is more likely to be synthesized. The $\Delta$G$_H$ values suggest  the highest HER performance for B@N1, Mn@N1 and Co@N1 configurations, owing to their moderate H-adsorption with |$\Delta$G$_H$| values less than 0.09 eV and may have catalytic activity surpassing that of Pt (111). These findings demonstrate the vital role of a coordinated environment as well as the doping configuration in tuning HER activity of g-C$_3$N$_4$. The ML models demonstrate that the SVR approach processed through FESF after hyperparameter tuning was the most suitable model as it has a stable and reliable predictive performance for the test set with a high R$^2$ of 0.81 and low MAE of 0.45. Based on the analysis of SVR model with feature engineering through statistical function, the result shows that the formation energy, bond length, boiling point, melting point, and valance electron are the key descriptor that influences the HER activity of SACs@g-C$_3$N$_4$. Overall, the present work not only provides a novel DFT-ML combined methodology that can efficiently and accurately screen HER catalyst but also reveals the significant features that determine the performance of g-C$_3$N$_4$-based single-atom catalysts.

\section{SUPPORTING INFORMATION}
Supporting information file contains PDOS, List of primary features or descriptors, Calculated vs predicted ∆G$_H$ and Hyperparameters.

\section{CONFLICTS OF INTEREST}
The authors declare no competing financial interest..

\section{DATA AVAILABILITY}
The data that supports the findings of this study are available within the article.

\section{ACKNOWLEDGEMENTS}
This work was supported by the Department of Science and Technology, Government of India, under the grant number SPO/DST/CHE/2021535. BMA would like to thank SERB for the financial support under the grant number PDF/2021/000487. We acknowledge the National Supercomputing Mission (NSM) for providing computing resources of ‘PARAM Sanganak’ at IIT Kanpur. Authors would like to thank D. H. K. Murthy, Manipal Institute of Technology for his valuable discussions and suggestions.

\end{document}